\documentclass[11pt]{article}
\usepackage[utf8]{inputenc}
\usepackage[dvipsnames]{xcolor}
\usepackage{algorithm2e}
\usepackage{enumitem}
\usepackage{url}
\usepackage{amsmath}%
\usepackage{amsfonts}%
\usepackage{amssymb}%
\usepackage{bm}%
\usepackage{graphicx}
\usepackage{cite}
\usepackage{amsthm}%
\usepackage{mathrsfs}
\usepackage{graphicx}
\usepackage{caption}
\usepackage{subcaption}
\usepackage{mathtools}
\usepackage{color}
\usepackage{xcolor}

\usepackage{mathtools}
\usepackage{bbm}
\usepackage{geometry}
\usepackage{setspace}
\usepackage{graphicx}
\usepackage{float}
\usepackage{color}
\usepackage{import}

\theoremstyle{definition}
\newtheorem{theorem}{Theorem}

\newcounter{assumption_}
\newtheorem{assumption}[assumption_]{Assumption}

\newcounter{remark_}
\theoremstyle{remark}
\newtheorem{remark}[remark_]{Remark}

\definecolor{myblue}{rgb}{0.6666,0.8,1}
\definecolor{myyellow}{rgb}{1,0.933,0.666}

\newcommand{\hinf}{\mathcal{H}_\infty}

\title{\bf Beyond KL-divergence: Risk Aware Control Through Cross Entropy and Adversarial Entropy Regularization}
\date{}

\author{Menno van Zutphen, Domagoj Herceg, Duarte J. Antunes\textit{}
\thanks{This research is part of the research program SYNERGIA (project number 17626), which is partly financed by the Dutch Research Council (NWO).}
\thanks{The authors are with the Control Systems Technology Group, Dep. of Mechanical Eng., Eindhoven University of Technology, the Netherlands. {email:\tt\small \{m.j.t.c.v.zutphen,\ d.herceg,\ d.antunes\}@tue.nl}. }} 

\begin{document}

\maketitle
\thispagestyle{empty}
\pagestyle{empty}
\begin{abstract} While the idea of robust dynamic programming (DP) is compelling for systems affected by uncertainty, addressing worst-case disturbances generally results in excessive conservatism. This paper introduces a method for constructing control policies robust to adversarial disturbance distributions that relate to a provided empirical distribution. The character of the adversary is shaped by a regularization term comprising a weighted sum of (i) the cross-entropy between the empirical and the adversarial distributions, and (ii) the entropy of the adversarial distribution itself. The regularization weights are interpreted as the \textit{likelihood factor} and the \textit{temperature} respectively. The proposed framework leads to an efficient DP-like  algorithm --- referred to as the \emph{minsoftmax} algorithm --- to obtain the optimal control policy, where the disturbances follow an analytical softmax distribution in terms of the empirical distribution, temperature, and likelihood factor. It admits a number of control-theoretic interpretations and can thus be understood as a flexible tool for integrating complementary features of related control frameworks. In particular, in the linear model quadratic cost setting, with a Gaussian empirical distribution, we draw connections to the well-known $\hinf$-control. We illustrate our results through a numerical example.
\end{abstract}



\vspace{-5mm}

\section{Introduction}

The system parameters and environmental conditions in real-world control applications are often subject to a significant level of uncertainty. Classical dynamic programming and other optimal control frameworks can exhibit a notable loss of performance in the presence of model mismatch. As a result, the need to appropriately handle the effect of uncertainty has motivated the design of \emph{robust} controllers~\cite{Ben-Tal:RobustOptimization}. Several practical approaches to robustness in control have been established over the years, including the classic $\hinf$-control \cite{Doyle1988StatespaceST}, minimax control (MM) \cite{Bertsekas1995DynamicPA}, risk-sensitive control~\cite{Howard1972RiskSensitiveMD} and the more recent distributionally robust control (DRC) methods \cite{robust-control-MDP-el-ghaoui-2005,Wiesemann2013RobustMD}. These methods have found application in a wide range of domains such as finance~\cite{follmer2011stochastic}, machine learning~\cite{hastie2009elements}, control~\cite{mesbah2016stochastic-overview} and others.

The minimax control framework designs a controller that minimizes cost w.r.t.\ the very worst-case disturbances. This often means that the controller protects the system against disturbances that are extremely unlikely to be encountered in practice. This extreme conservatism generally results in poor performance under close to nominal conditions. The cornerstone method of robust control, $\hinf$-control~\cite{basar2008_hinf_minimax}, is able to somewhat reduce this conservatism by regularizing the energy at the disposal of the adversary playing against the controller. While this is often a significant improvement over minimax control, the energy of a disturbance is generally not well-defined in finite spaces and only reflects the likelihood of the disturbances in specific continuous space scenarios such as the Gaussian setting. Distributionally robust control~\cite{DRO_frameworks} (DRC) instead assumes an \emph{ambiguity} set within which the actual underlying probability distribution of the disturbance is contained. DRC methods then design a controller that is guaranteed a certain level of performance w.r.t.\ all the distributions in the uncertainty set. Risk-Sensitive control penalizes higher-order moments, such as the variance of the cost. By disproportionately considering high-cost outcomes, the method can be interpreted as robust against variations of the disturbance distribution. 

\begin{figure}[t]
    \centering
    \input{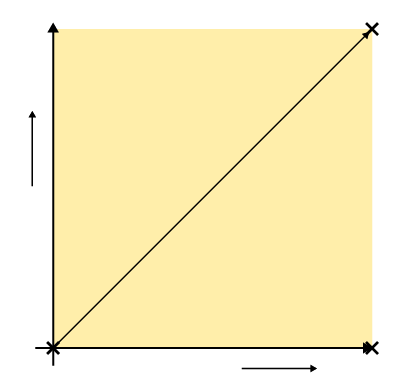}
    \caption{By selecting $(\gamma_{H},\gamma_{E})$, we can traverse the design space of proposed \emph{minsoftmax} controllers. Selecting a policy that is close to 
 (and in the limit boils down to), e.g., $\hinf$-control, minimax control (MM), stochastic dynamic programming (SDP, see Remark \ref{rem:SDP}), maximum likelihood certainty equivalence (ML-CE, see Remark \ref{rem:ml-ce}), or risk-sensitive control (see Remark \ref{rem:risk-sensitive}), but also inherits interesting features from the others.}
    \label{fig:map}
\end{figure}

This paper introduces a method for constructing risk-aware control policies that take into account adversarial disturbance distributions, by penalizing deviations from an empirical one. A similar method was presented in~\cite{minimax_wasserstein_DR} for the Wasserstein distance penalty. Specifically, we propose adding a novel regularization term to the minimax framework, composed of a weighted sum of (i) the cross-entropy between the empirical and the adversarial distributions, and (ii) the entropy of the adversarial distribution itself.\footnote{Similar techniques have been used in reinforcement learning for policy regularization~\cite{soft-actor-critic,LeverageTheAverage}.} The corresponding regularization weights are interpreted as the \textit{likelihood factor} $\gamma_{H}$, which steers the considered adversarial disturbance distributions away from highly unlikely disturbances, and the \textit{temperature} $\gamma_{E}$, encouraging similarity to the empirical. As a result, as Fig.~\ref{fig:map} suggests, our method can be seen as a flexible tool for integrating complementary features of well-known control frameworks, such as $\hinf$-control, stochastic DP, minimax control, certainty-equivalent control, and KL-regularized control.

We prove that our proposed framework leads to an efficient DP-like  algorithm --- referred to as the \emph{minsoftmax} algorithm --- to obtain an optimal control policy, where disturbances follow an analytical softmax distribution in terms of the empirical distribution, temperature, and likelihood factor. In addition, we draw connections to the well-known $\hinf$-control frameworks~\cite{basar2008_hinf_minimax}. Specifically, an optimal control policy for the proposed framework, when the model is linear, the cost is quadratic, and \textit{the empirical distribution is Gaussian} is shown to coincide with a well-known optimal policy from $\hinf$-control. Finally, we illustrate the benefits of our method through numerical examples .


\def\svgwidth{0.45\textwidth} 


\section{ Problem formulation}\label{sec:2}

\textit{Notation:}  The set of all probability density functions defined over $\mathbb{R}^{n}$ is denoted by $\mathcal{P}^{n}:=\{p : \mathbb{R}^{n}\rightarrow\mathbb{R}_{\ge 0} \mid \int_{\mathbb{R}^{n}}p(x)~\mathrm{d}x=1\}$. The set of all probability distributions over a finite alphabet of cardinality $n\in\mathbb{N}$, is denoted by $\Delta^{n}:=\{p\in\mathbb{R}^{n}_{\ge 0}\mid\sum_{i}p_i=1\}$, i.e., the $(n-1)$-dimensional simplex.

Consider a discrete-time control system
\begin{equation}\label{eq:sys}
	x_{k+1}=f(x_k,u_k,w_k),
\end{equation} 
where $x_k \in \mathcal{X}$, $u_k\in \mathcal{U}$, $w_k \in \mathcal{W}$ are the state, the control input and the disturbance at time $k \in \mathcal{K}$, $\mathcal{K}:=\{0,1,\dots,h-1\}$.  
 We consider both the distinct cases where $\mathcal{X}=
 \mathbb{R}^n$, $\mathcal{U}=\mathbb{R}^{n_{\text{u}}}$, $\mathcal{W}=\mathbb{R}^{n_{\text{w}}}$ are continuous (Euclidean) and where $\mathcal{X}=
 \{1,2,\dots, \mathsf{n}
 \}$, $\mathcal{U}=\{1,2,\dots, \mathsf{n_u}
 \}$, $\mathcal{W}=\{1,2,\dots, \mathsf{n_w}
 \}$ are discrete (finite). Whenever we state results that apply to both continuous and discrete space systems, we will use continuous state notation as the discrete alternative is analogous in a straightforward way.  The disturbances $w_k$, $k \in \mathcal{K}$, are assumed to be independent random variables whose distribution can depend on the state and input at time $k$. When considering continuous spaces, we will assume absolute continuity of the probability measure governing each disturbance $w_k$ with respect to the Lebesgue measure, ensuring the availability of a probability density function representation. Such a probability density function (conditioned on the state and input at time $k$) is denoted by $p_k(\cdot |x_k,u_k)\in\mathcal{P}^{n_{\text{w}}}$, i.e., $\text{Prob}[w_k 
 \in \mathcal{A}|x_k=\mathsf{x},u_k=\mathsf{u}]=
 \int_{\mathcal{A}}p_k(w|\mathsf{x},\mathsf{u})\mathrm{d}w$, for a measurable set $\mathcal{A}\in \mathbb{R}^{n_{\text{w}}}$.  We assume that some knowledge about these probability distributions is available, namely through an empirical distribution $r_{k}(\cdot |
 x_k,u_k)\in\mathcal{P}^{n_{\text{w}}}$, that can be seen as an estimate of $p_k(\cdot |x_k,u_k)$. The empirical distribution is also assumed to be absolutely continuous. This empirical distribution can be obtained by, e.g., fitting a class of absolutely continuous distributions to data. For the sake of compactness we use $w \sim p$ (and $w \sim r$) to indicate that each $w_k$ is distributed according to $p_k(\cdot |x_k,u_k)$ (and $r_{k}(\cdot |x_k,u_k)$, respectively).

In traditional stochastic control, the aim is to find an input policy $\mu=(\mu_0,\mu_1,\dots,\mu_{h-1})$, such that $u_k=\mu_k(x_k)$ minimizes an expected cumulative cost w.r.t. a given disturbance (empirical) distribution 
\[
\inf_{\mu}\underset{w\sim r}{\mathbb{E}}\left[\sum_{k=0}^{h-1} g(x_k,\mu_k(x_k))+g_h(x_h)\right].
\]
While minimizing over control policies we will consider the infimum instead of the minimum, as a minimizing policy might not exist in uncountable spaces. Even then, a control policy can be found that results in a cost arbitrarily close to the infimum (provided the infimum is not at $-\infty$). 
In practical scenarios, the values available to describe the empirical disturbance distribution might come in the form of an estimate based on data, or as a basis distribution describing a large set of systems. To mitigate the resulting uncertainty around a (nominal) disturbance distribution, one might wish to design a controller that achieves a certain level of performance w.r.t. any of the possible true underlying distributions. This scenario gives rise to the problem of \emph{distributionally robust control} if  the empirical distribution is restricted to lie in some set, and \emph{minimax} control in case no such information is available and all possible distributions are considered. In this paper, we propose an alternative, which we formulate as follows. 

Let
\[
    H_c(p_k,r_k)=-\int_{\mathcal{W}}p_k(w|x_k,u_k)\log r_k(w|x_k,u_k)\mathrm{d}w,
\]
denote the cross-entropy of $p_k(\cdot |x_k,u_k)$ w.r.t.\ $r_k(\cdot |x_k,u_k)$ and
\[
    H(p_k)=-\int_{\mathcal{W}}p_k(w|x_k,u_k)\log p_k(w|x_k,u_k)\mathrm{d}w,
\]
denote the entropy of $p_k(\cdot |x_k,u_k)$. We then propose to tackle the following problem
\begin{equation}
    \inf_{\mu}J_{\mu}(x_0,\gamma_{H},\gamma_{E}), \label{eq:minsoftmaxproblem}
\end{equation}
where, for some positive constants $\gamma_E$ and $\gamma_H$, 
\begin{equation}
\begin{split}
    J_{\mu}(x_0,\gamma_{H},\gamma_{E})=\hspace{20mm}&\\
    \sup_{p}\underset{w\sim p}{\mathbb{E}}\bigg[\sum_{k=0}^{h-1} g(x_k,\mu_k(x_k))- &\gamma_H H_{\text{c}}(p_k,r) \\
    &+\gamma_E H(p_k)+g_h(x_h)\bigg].
\end{split}\label{eq:minsoftmaxcost} 
\end{equation}
Since we maximize over disturbances in~\eqref{eq:minsoftmaxcost}, we call $p_k(\cdot |x_k,u_k)$ the adversarial disturbance. 

This formulation~\eqref{eq:minsoftmaxcost} has its roots in KL-ball distributionally robust control~\cite{hu2013kullback}. KL-ball DRC constrains the adversary to pick exclusively from distributions $p_k(\cdot |x_k,u_k)$ that have a Kullback-Leibler divergence
\[
\operatorname{KL}(p_k\|r_k)=\int_{\mathcal{W}}p_k(w|x_k,u_k)\log \frac{p_k(w|x_k,u_k)}{r_k(w|x_k,u_k)}~\mathrm{d}w,
\]
smaller than some $\varepsilon\in\mathbb{R}_{\ge 0}$ w.r.t.\ the empirical distribution
\[
\mathcal{B}(r_k,\varepsilon):=\{p_k\in\mathcal{P}^{n}\mid \operatorname{KL}(p_k\|r_k)\le \varepsilon\}.
\]
It is well known that the inner (adversary) problem arising from the KL-ball DRC setup can be solved as a line search over a Lagrange variable that essentially regulates the weight of a \textit{KL-divergence penalty} \cite{robust-control-MDP-el-ghaoui-2005}. Moreover, we have the following key identity:
\begin{equation}
    \operatorname{KL}(p_k||r_k) = H_{\text{c}}(p_k, r_k)-H(p_k). \label{eq:KL_identity}
\end{equation}

\par We can then interpret the proposed problem formulation as including an additional degree of freedom in this \emph{soft-constrained} variant of the DRC. In fact, the intended KL-divergence regularization term and its individual (cross-entropy and entropy) terms are weighted individually, yielding~\eqref{eq:minsoftmaxcost}, which, as we will show in this paper, turns out to provide a number of attractive properties. This newly proposed cost and its associated control problem
will serve as the basis of the analysis in this paper, in which we show that the formulation naturally gives rise to a computationally attractive solution algorithm and outline its many control-relevant properties. 

\begin{remark}
Due to the recursive nature of cross-entropy and entropy when applied to trajectory probabilities in Markov systems, cost~\eqref{eq:minsoftmaxcost} can alternatively be interpreted as
\[
\begin{split}
\sup_{p}\underset{w\sim p}{\mathbb{E}}\Big[\sum_{k=0}^{h-1}& g(x_k,u_k)+g_h(x_h)\Big] \\
&- \gamma_H H_{\text{c}}(T_p,T_r)+\gamma_E H(T_p),
\end{split}
\]
with the cross-entropy and entropy as defined above, and $T_p$ and $T_r$ the joint probability density functions of the disturbances, i.e., 
\[
\begin{split}
T_r(w_0,\cdots, w_{h-1} ):=\hspace{20mm}& \\
\operatorname{Prob}(w_0, \!\cdots, w_{h-1}  \mid 
x_{k+1}&=f(x_k,u_k,w_k), \\ u_k&=\mu_k(x_k), \\
w_k&\sim r_k(\cdot | x_k,u_k)),
\end{split}
\]
and $T_p$ is defined similarly. This amounts to directly regularizing the original cost by the cross-entropy and entropy \emph{of the entire disturbance trajectory}.
\end{remark}

\color{black}

\color{black}

\section{Methods and results}\label{sec:3}

In this section, we discuss our findings regarding problem (\ref{eq:minsoftmaxproblem}), starting with the observation that our inner problem, i.e. the search for adversarial disturbances in~\eqref{eq:minsoftmaxcost},  admits a closed-form solution.

\begin{theorem}\label{thm:analytical_solution_continuous}
Consider problem~\eqref{eq:minsoftmaxproblem}, \eqref{eq:minsoftmaxcost}. Let $J_h(x) = g_h(x)$, then consider the following recursion
\begin{equation}
\begin{split}
    &J_k(x)=\inf_{u\in\mathcal{U}}\sup_{p\in\mathcal{P}^{n_{\text{w}}}}g(x,u) + -\gamma_{H}H_{\text{c}}(p_k,r)\\
    &\hspace{2cm}+\gamma_{E}H(p_k) + \underset{w\sim p_k}{\mathbb{E}}J_{k+1}(f(x,u,w)),\label{eq:dp_solution_original}
\end{split}
\end{equation}
for $k\in \{h-1,h-2,\dots,0\}$. We find that an optimal adversary of~\eqref{eq:dp_solution_original}, for every $(x,u,k)\in\mathcal{X}\times\mathcal{U}\times\mathcal{K}$, can be described in closed form as
\begin{equation}
    p_k^{*}(w|x,u) = \frac{e^{\alpha_{k}(x,u,w)/\gamma_E}}{\int_{\hat{w}\in \mathcal{W}}e^{\alpha_{k}(x,u,\hat{w})/\gamma_E}~\mathrm{d}\hat{w}},\label{eq:softmax_optimal_adversary}
\end{equation}
where
\begin{equation}
\begin{split}
    \alpha_k(x,u,w) &= \gamma_{H}\log r(w |
    x,u)+ J_{k+1}(f(x,u,w)), \label{eq:alpha}
\end{split}
\end{equation}
which, when substituted together into~\eqref{eq:dp_solution_original}, yields the equivalent cost
\begin{align}
    J_k(x) = \inf_{u\in \mathcal{U}} g(x,u)+Q_k(x,u), \label{eq:value_iteration_main}
\end{align}
where
\begin{equation}
Q_k(x,u)=\begin{cases}
    \gamma_{E} \log \underset{\mathcal{W}}{\int}e^{\alpha_k(x,u,w)/\gamma_{E}}~\mathrm{d}w, &\text{for } \gamma_{E}>0, \\
    \sup_{w\in \mathcal{W}} \ \alpha_k(x,u,w), & \text{for } \gamma_{E}=0.
    \end{cases}\label{eq:D3}
\end{equation}
Then, if a minimizer exists in~\eqref{eq:value_iteration_main} for every $x\in \mathcal{X}$, and every $k\in \mathcal{K}$, an optimal control policy for~\eqref{eq:minsoftmaxproblem},~\eqref{eq:minsoftmaxcost} is given by 
\begin{align}
    \mu_k(x)& \in \arg \inf_{u\in\mathcal{U}}\ g(x,u)+Q_k(x,u).\label{eq:mu}
\end{align}
\begin{flushright} $\square$ \end{flushright}
\end{theorem}
\begin{proof}
See Section \ref{sec:technical_results}.
\end{proof}

The key feature that enables us to obtain the optimal policy with this theorem is that at each time $k$, we obtain an explicit expression for the adversarial disturbance policy in terms of a softmax function (we borrow the term from discrete space setting), such that we can compute the minimum (when it exists) over control decisions. For this reason, we call the DP-like algorithm in Theorem~\ref{thm:analytical_solution_continuous} the \emph{minsoftmax} algorithm.

\subsection{Finite spaces}

Note that the results described in Theorem~\ref{thm:analytical_solution_continuous} hold for finite spaces, after standard modifications such as interpreting the integrals as summations. As a minimizer for~\eqref{eq:value_iteration_main} always exists in finite spaces, an optimal control policy for~\eqref{eq:minsoftmaxproblem} can be computed using the simple dynamic programming value iteration Algorithm~\ref{alg:1}. This approach is obtained by substituting the identities found in Theorem \ref{thm:analytical_solution_continuous} into the recursion obtained by decomposing the cumulative cost (\ref{eq:minsoftmaxcost}). Implcit to the computation of $J_k(x_k)$, the softmax adversarial policy, $p(w|x_k,u_k) = \frac{e^{\alpha_w/\gamma_E}}{\sum_{\hat{w}=1}^{n_{\text{w}}}e^{\alpha_{\hat{w}}/\gamma_E}}$, is selected, where $\alpha_w = \gamma_H \log r(w|x_k,u_k)+J_{k+1}(f(x_k,u_k,w))$, for $w\in\{1,2,\dots,n_{\text{w}}\}$. It is this property that leads us to refer to the algorithm as the \emph{minsoftmax algorithm}.

\begin{algorithm}[h!]
\hrule\vspace{2pt} 
\hrule\vspace{2pt} 
\caption{ Minsoftmax control in finite spaces}\label{alg:1}
Set $J_h(x) = g_h(x)$.

\For{ $k \in \{h-1,h-2,\dots,0\}$ }
    {
    \For{$x \in \{1,2,\dots,|\mathcal{X}|\}$}
        {
        \For{$u \in \{1,2,\dots,|\mathcal{U}|\}$}
            {
            \For{$w\in\{1,2,\dots,n_{\text{w}}\}$}
                {
                \vspace{-2mm}
                
                \[
                \begin{split}
                &\alpha_w = \gamma_H \log r(w|x,u)\\
                &\hspace{2cm} + J_{k+1}(f(x,u,w)),
                \end{split}
                \]
                \vspace{-4mm}
                }
            \vspace{-5mm}
            \begin{equation}
            \hspace{-4mm}Q(u) = \begin{cases}
                \gamma_{E}\log\underset{w\in \mathcal{W}}{\sum}e^{\frac{\alpha_w}{\gamma_{E}}} &\text{if }\gamma_{E}>0, \\
                \max_{w \in \mathcal{W}} \alpha_{w}, &\text{if } \gamma_{E}=0, 
            \end{cases}\label{eq:algo_Qu}
            \end{equation}
            }
        } 
        \vspace{-6mm}
        \begin{equation}
        \begin{split}
        \mu_k(x)& = \arg \min_{u\in\mathcal{U}}g(x,u)+Q(u),\label{eq:mu}\\
        J_k(x) &= g(x,\mu_k(x))+Q(\mu_k(x)).
        \end{split}
        \end{equation}
        \vspace{-4mm}
    }
\hrule\vspace{2pt} 
\hrule\vspace{2pt} 
\end{algorithm}

\subsection{Tuning the penalty parameters}

The setting $\gamma_{E}=\gamma_{H}=0$ is straightforwardly interpreted as unregularized minimax worst-case minimization, see (\ref{eq:minsoftmaxcost}). As this yields the most robust policy, an increase in the values of $\gamma_{H}$ and $\gamma_{E}$ is generally desired to reduce conservatism. Below, we discuss interpretations of the increases in these penalization weights.

We interpret $\gamma_{E}$ as the softmax \emph{temperature} of (\ref{eq:softmax_optimal_adversary}), as increasing it has the effect of randomizing the adversary. The interpretation of $\gamma_{H}$ is best understood when considering $\gamma_{E}=0$. When $\gamma_{H}=0$, the adversary will simply select the disturbance realization associated with the worst cost, see~\eqref{eq:D3}. Instead, after raising the value of $\gamma_{H}$, the adversary is penalized for selecting highly unlikely disturbances. Parameter $\gamma_{H}$ is thus dubbed the \emph{likelihood factor}, as it encourages the adversary to select increasingly likely disturbances. 

\begin{remark}[Recovering maximum likelihood certainty equivalence control]\label{rem:ml-ce}
    We note that the likelihood interpretation, when taken to the extreme $\gamma_{E}=0$, $\gamma_{H}\to\infty$, recovers the maximum likelihood certainty equivalence control policy \cite{Cai2021ASB}. This can be confirmed by subtracting the constant term $\gamma_{H}\log\max_{\hat{w}\in\mathcal{W}} r(\hat{w})$ from~\eqref{eq:alpha} and taking the limit $\gamma_{H}\to\infty$. This makes any adversary choice outside of $w^{*}=\arg\max_{\hat{w}\in\mathcal{W}}r(\hat{w})$ evaluate to $-\infty$, while choice $w^{*}$ yields cost $J_{k+1}(f(x,u,w^{*}))$, recovering maximum-likelihood certainty equivalence.
\end{remark}

For all $\gamma_{E}=\gamma_{H}$, we recover KL-divergence regularized cost (see~\eqref{eq:KL_identity}). Algorithm \ref{alg:1} can be seen to simplify under these conditions, as $\gamma_{H}/\gamma_{E}=1$, to match the \emph{risk-sensitive} control solution.

\begin{remark}[Recovering risk-sensitive control]\label{rem:risk-sensitive}
Risk-sensitive control comprises the problem of minimizing risk-sensitive cost
\[
\min_{\mu} \underset{w\sim r}{\mathbb{E}}\left[\exp(\gamma \sum_{k=0}^{h-1} g(x_k, u_k))\right],
\]
where $\gamma\in\mathbb{R}$ manages the sensitivity to risk ($\gamma<0$: risk-seeking, $\gamma \searrow 0$: risk-neutral, $\gamma>0$: risk-averse). The log-transformed risk-sensitive dynamic programming algorithm iterates 
\[
\min_{u}g(x, u)+\frac{1}{\gamma} \log\underset{w\sim r}{\mathbb{E}} e^{\gamma J(f(x,u,w))},
\]
which can be seen to coincide with our solution~\eqref{eq:mu}, when substituting in \eqref{eq:D3}, with \eqref{eq:alpha}, for $0<\gamma_{E}=\gamma_{H}=\frac{1}{\gamma}$.  
\end{remark}

Lastly, taking the limit $\gamma_{E}=\gamma_{H}\to\infty$, our adversary is pushed towards the empirical distribution.

\begin{remark}[Recovering stochastic dynamic programming]\label{rem:SDP}
As KL-divergence is always positive except when its arguments are equal, and $\gamma_{E}=\gamma_{H}$ reduces the regularization term to the negative KL-divergence between $p$ and $r$, increasing this weight has the effect of leaving only a single choice of $p$ with non-negative infinite cost. The optimal adversary thus becomes: $p^{*}=r$.
\end{remark}

Finally, by varying the value of the parameters inside the region $0<\gamma_{E}\le\gamma_{H}<\infty$, the extent to which the features of the aforementioned control paradigms show up in our robust controller can be selected, and the controller can be made more/less robust to unlikely disturbances $0<\gamma_{H}$, and disturbances that behave unlike the empirical distribution $0<\gamma_{E}\le\gamma_{H}$.

\subsection{General spaces and the connection to $\hinf$-control}\label{sec:general_spaces}

To study the case of continuous spaces, we impose some simplifying assumptions, namely that the model is linear and the cost is quadratic. In addition, to address infinite horizons, we impose standard observability and controllability assumptions.

\begin{assumption}\label{as:1} \ \
Well-posedness
    \begin{enumerate}[label=(\roman*)]
    \item $f(x,u,w) = Ax+Bu+Dw$.
    \item $g(x,u) = x^{\top} Q x+u^{\top} R u$, $g_h(x) = x^{\top} Q_h x$.
    \item $Q_h\succeq$, $Q\succeq 0$, $R\succ 0$.
    \item $(A,Q)$ detectable, $(A,B)$ controllable.
    \end{enumerate}
\end{assumption}

The following assumption that imposes a Gaussian empirical distribution as a prior on our adversary is key to making a connection between problem~\eqref{eq:minsoftmaxproblem},~\eqref{eq:minsoftmaxcost}, and $\hinf$-control.
\begin{assumption}\label{as:2}  $r_k(\cdot |x,u)$ is zero-mean Gaussian with identity covariance, denoted by $\mathcal{N}(0,I)$, for every $x \in \mathbb{R}^n$, $u \in \mathbb{R}^{n_{\text{u}}}$, and $k\in\mathcal{K}$.
\end{assumption}

Note that this assumption is less limiting than it might appear, as the character of more complex distributions can often be absorbed into $D$ to recover identity covariance. Suppose that $\gamma_E=0$. When the empirical distribution satisfies Assumption~\ref{as:2}, the maximization in the right-hand side in~\eqref{eq:D3} is equivalent to the following simple maximization 
\[\sup_{w\in \mathbb{R}^{n_{\text{w}}}}-\gamma_H\frac{1}{2}w^{\top} w+J_{k+1}(f(x,u,w)).
\]
The crucial fact to note here is that this disturbance policy is deterministic. We can therefore reconsider the problem~\eqref{eq:minsoftmaxproblem},~\eqref{eq:minsoftmaxcost} by restricting the class of adversarial disturbances to be deterministic, denoted by $\eta = (\eta_0,\eta_1,\dots,\eta_{h-1})$ such that $w_k=\eta_k(x_k,u_k)$; note that the disturbances in this formulation are allowed to depend on the control input, as we consider only the original min-max problem (and never its reverse), where the minimization is taken with respect to the control policy and the maximization with respect to the disturbance policy. This new deterministic disturbance policy problem boils down exactly to the soft-constrained linear-quadratic dynaics game~\cite[Ch. 3]{basar2008_hinf_minimax} (that leads to $\hinf$ control as explained in the sequel), that is, 
\begin{equation}
\begin{split}
\inf_{\mu}
\sup_{\eta}\underset{w\sim \eta}{\mathbb{E}}\bigg[\sum_{k=0}^{h-1} g(x_k,u_k)- &
\gamma^2w_k
^{\top} w_k+g_h(x_h)\bigg].
\end{split}\label{eq:minsoftmaxcost2} 
\end{equation}
with $u_k = \mu_k(x_k)$, $w_k = \eta_k(x_k)$, for every $k\in \mathcal{K}$,

 \begin{equation}\label{eq:D6} \gamma^2 = \frac{\gamma_H}{2}.\end{equation}
However, somewhat surprisingly, the same policy is obtained for $\gamma_E>0$, as stated in the next result.

\begin{theorem}\label{thm:continous_recursion}
Suppose that Assumptions~\ref{as:1},~\ref{as:2} are satisfied. 
Then the cost functions for the dynamic programming algorithm in Theorem~\ref{thm:analytical_solution_continuous} are given by
\begin{align}
J_k(x_k) = x_k^\top P_k x_k + \zeta_k,
\end{align}
where the $P_k$, $k \in \mathcal{K}$, can be computed by the following recursion with $P_h = Q_h, \zeta_h = 0$. 
For $k \in \{h-1,h-2,\dots,0\}$ iterate
\begin{align}
P_{k} &=F_c(F_a(P_{k+1})),
\end{align}
where 
\[
\begin{aligned}
    F_a(P)&:= P+PD(\gamma_H I-2D^\top PD)^{-1}D^\top P, \\
    F_c(P)&:= Q + A^\top PA-A^\top PB(B^\top PB+R)^{-1}B^\top PA,
\end{aligned}
\]
provided that $\gamma_H$ is such that 
\begin{equation}\label{eq:Mk}
M_k = \gamma_H I - 2D^\top P_k D \succ 0,
\end{equation}
for every $k \in \{1,\dots,h\}$. 

The additive cost offset $\zeta_k$ is found as
\begin{equation}
\begin{split}
&\zeta_k = (\gamma_{E}-\gamma_{H})\log (2\pi)^{n_{\text{w}}}/2\\
&\hspace{2cm}-\gamma_{E}\log(\det(M_{k+1})/\gamma_{E}^{n_{\text{w}}})/2 + \zeta_{k+1}.
\end{split}\label{eq:zeta_update}
\end{equation}
for $k \in \{h-1,h-2,\dots,0\}$.

The optimal control policy for~\eqref{eq:minsoftmaxproblem},~\eqref{eq:minsoftmaxcost} is the following linear control law 
\begin{equation}
\label{eq:1} 
u_k^* =  -G(P_{k+1})x_k,
\end{equation}
where 
\begin{align}
G(P)=(R+B^\top F_a(P)B)^{-1}B^\top F_a(P)A.
\end{align}
Moreover, an optimal adversarial distribution is a Gaussian with state-dependent mean, and modified covariance which scales linearly with the temperature parameter $\gamma_E$ and is given by
\begin{align}\label{eq:F13}    
p_k^*(x_k) = \mathcal{N} \left(M_{k+1}^{-1} 2 D P_{k+1} A_{G_k} x_k , \gamma_E M_{k+1}^{-1} \right) 
\end{align}
where $A_{G_k} = A - BG(P_{k+1})$.
\end{theorem}
\begin{proof}
See Section \ref{sec:technical_results}.
\end{proof}

\par Notice that the above iteration for $P_k$ is not a function of $\gamma_E$. In Section~\ref{sec:5}, different policies were obtained as a result of a change in $\gamma_E$ for a fixed $\gamma_H$. However, this turns out to not be the case under Assumptions~\ref{as:1},~\ref{as:2}.
\par One can conclude from Theorem~\ref{thm:continous_recursion} that the mean of the adversarial disturbance policy~\eqref{eq:F13} coincides with a worst-case disturbance policy for the soft-constrained linear-quadratic dynamic game considered in~\cite[Ch. 3]{basar2008_hinf_minimax}. \par As $h\rightarrow \infty$, and provided that~\eqref{eq:Mk} holds for every $k$, $K_k \rightarrow K$, $K = G(\bar{P})$ where $\bar{P}$ is the unique positive semi-definite solution to $\bar{P}=F_c(F_a(\bar{P}))$,
and the resulting policy $u_k = -G(P) x_k$, coincides with that of $\hinf$ control~\cite{basar2008_hinf_minimax}. This policy guarantees that the following cost is negative 
 \begin{equation}\label{eq:costgame}
 	\inf_{\mu} \sup_{\eta} \sum_{k=0}^\infty x_k^{\top} Q x_k-\gamma^2 w_k^{\top} w_k,
 \end{equation}
which implies that 
\[
\inf_{u_k=\mu(x_k)} \sup_{w\in \ell_2} \frac{\sum_{k=0}^\infty z_k^{\top} z_k}{\sum_{k=0}^\infty w_k^{\top} w_k}\leq \gamma^2,
\]
where $z_k = Q^{1/2}x_k$. \color{black}The smallest attenuation bound is obtained by minimizing $\gamma$ for which $\gamma^2I>P_k$ for every $k$. 
Therefore, by considering a horizon converging to $\infty$, the gains of the policy we obtain converge to those of $\hinf$-control that guarantees a given $\ell_2$ induced gain. 

Taking into account the considerations made pertaining to Fig.~\ref{fig:map}, we can also conclude the following:
\begin{itemize}
    \item[(i)] When $\gamma_E=0$, as $\gamma_H \rightarrow \infty$  we obtain certainty equivalent control, which for the linear quadratic case boils down to
    \begin{equation}\label{eq:LQRpolicy} 
        u_k = L_k x_k,
    \end{equation}
    for 
    \[
    L_k = -(R+B^{\top} X_{k+1} B)^{-1}B^{\top} X_{k+1}A,
    \]
    where $X_h = Q_h$ and for $k\in \{h-1,h-2,\dots,0\},$ 
    \[ 
    \begin{split}
    &X_k = Q+B^{\top} P B\\
    &\hspace{1cm}+A^{\top} X_{k+1}B(R+B^{\top} X_{k+1} B)^{-1}B^{\top} X_{k+1}A.
    \end{split}
    \]
   \item[(ii)] When $\gamma_H=\gamma_E$, and both approach $\infty$, we obtain the optimal stochastic dynamic policy which coincides with~\eqref{eq:LQRpolicy}. 
    \item[(iii)] When both $\gamma_H \rightarrow 0$ and $\gamma_E \rightarrow 0$ we obtain min-max control and due to the power given to disturbances the cost becomes unbounded. Actually the cost becomes unbounded for a critical value of $\gamma_H$ (the infimum value such that~\eqref{eq:Mk} holds) and it is independent of 
    $\gamma_E$.
\end{itemize}


\color{black}
\begin{remark}[Interpretation: $\hinf$-control equivalent algorithm for discrete spaces]
Seeing that the $\gamma_{E}=0$ minsoftmax method for this class of continuous space problems yields the $\hinf$-controller and costs when stage-cost $g(x_k,u_k)=x_k^\top x_k$, we believe it is natural to denote the equivalent problem set-up in finite spaces as \emph{the finite space equivalent of $\hinf$-control}.
\end{remark}

\section{Numerical examples}\label{sec:5}

\begin{figure}[t]
    \centering
    \includegraphics[width=0.6\linewidth]{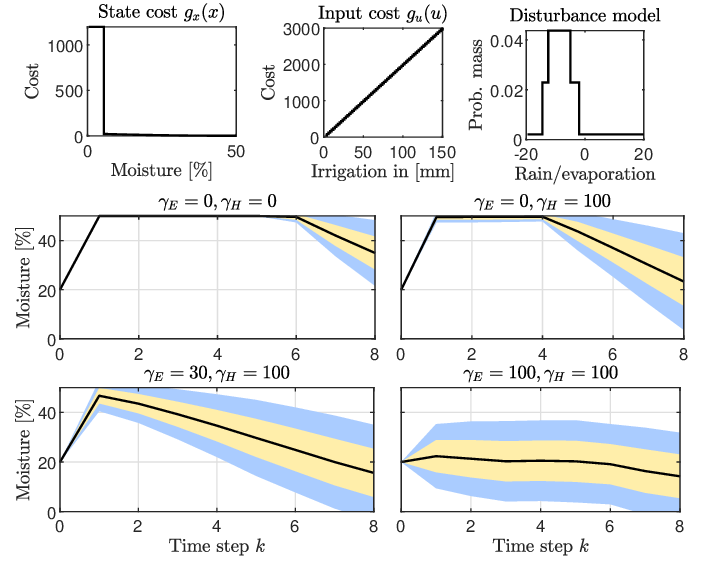}
    \caption*{
    \makebox{
    \begin{minipage}{0.5\linewidth}
    \textbf{-----} mean of simulated trajectories\\ 
    \fcolorbox{black}{myyellow}{\rule{0pt}{1ex}\rule{2ex}{0pt}} $\sigma$ region of simulated trajectories \\
    \fcolorbox{black}{myblue}{\rule{0pt}{1ex}\rule{2ex}{0pt}} $2\sigma$ region of simulated trajectories
    \end{minipage}}} 
    \caption{The simulated example setting for $h=8$ and $(\gamma_{E},\gamma_{H})\in[0,100]^{2}$.}
    \label{fig:simulation}
\end{figure}

\begin{figure}[t]
    \centering
    \includegraphics[width=0.6\linewidth]{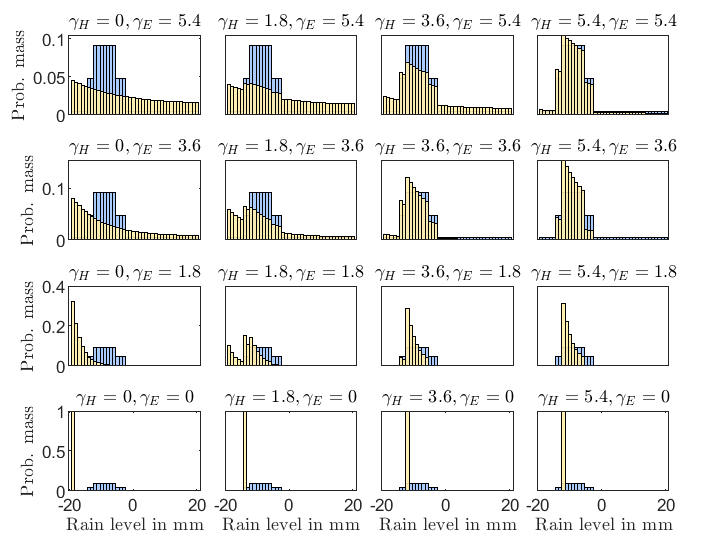}
    \caption*{\fcolorbox{black}{myblue}{\rule{0pt}{1ex}\rule{2ex}{0pt}} empirical disturbance distribution \\ 
    \hspace{2.8mm}\fcolorbox{black}{myyellow}{\rule{0pt}{1ex}\rule{2ex}{0pt}} adversarial disturbance distribution}
    \caption{Example of a single optimal adversarial distribution as a function of the empirical disturbance distribution $r$, different values $(\gamma_{E},\gamma_{H})\in[0,5.4]^2$ and the cost $g_h(x)=g_x(x)$, with $x=8$. Note that the top-left modes ($\gamma_{E}>\gamma_{H}$) are not considered of practical use.}
    \label{fig:adversary}
\end{figure}

\subsection{Simulations}

To illustrate the effect our method has when applied to dynamic systems, we consider an example from the field of agriculture. Consider a field that is to be irrigated. Its moisture content level $m\in[0,50]$ \%, is mapped to the discrete states $x\in\{1,2,\dots,100\}$. This moisture level is affected by evaporation/rain $e\in[-20,20]$ mm, modeled discretely by $w\in\{1,2,\dots,40\}$. The available empirical prediction model $r\in\Delta^{40}$ of evaporation/rain is available as a weighted average of three uniform distributions. A 100\% confidence interval over $[-20,20]$ mm, and two 95\% confidence intervals over $[-15,-5]$ and $[-13,-2]$, obtained, e.g., from separate weather stations, see Fig.~\ref{fig:simulation} (top right). To control the moisture content level, one is able to make irrigation decisions of $i\in[0,150]$ mm, mapped to the discrete $u\in\{1,2\dots,75\}$. The system then evolves as
\[
s_{k+1} = \max\{\min\{m(x_k) + i(u_k)/3 + e(w_k),50\},0\},
\]
where functions $m(x)$, $i(u)$, and $e(w)$ map the finite $x$, $u$, and $w$, to their corresponding continuous space values, and $x_{k+1}=\beta(s_{k+1})$ is used to map the predicted moisture level $s_{k+1}$ back to the nearest representative state in the finite domain. A separable stage cost over the states and inputs is considered, as
\[
g(x,u)=g_x(x)+g_u(u),
\]
where
\[
g_u(u)=20i(u), \ \ 
g_x(x) = \begin{cases}
    1200 & \text{if } x\in\{1,\dots,10\}, \\
    \frac{(s(x)-50)^2}{100} & \text{otherwise}.
\end{cases}
\]
As irrigation simply costs fuel and human resources, its cost scales linearly, while low moisture content scales quadratically in reduced plant yields and incurs a heavy cost when it dips below the level compatible with life. These cost functions are visualized in the top row of Fig.~\ref{fig:simulation}.

The performance of the minsoftmax controllers for each of the four variations of $(\gamma_{H},\gamma_{E})\in\{(0,0),(0,100),(30,100),(100,100)\}$ on the example system for $h=8$ has been simulated 5000 times w.r.t.\ the nominal disturbance model. The resulting mean and variance of these trajectories are displayed in Fig.~\ref{fig:simulation}.

From these resulting trajectory distributions, it becomes clear that the effect of reducing conservatism through the raising of just the value of the likelihood factor $\gamma_{H}$ can be limited. This can be explained by observing Fig.~\ref{fig:adversary}, where the $\gamma_{E}=0$ adversaries can be seen to always represent only a single likely bad disturbance, while discarding all other empirical distribution information.

\subsection{Minsoftmax design considerations}

\begin{figure}[t]
    \centering
    \input{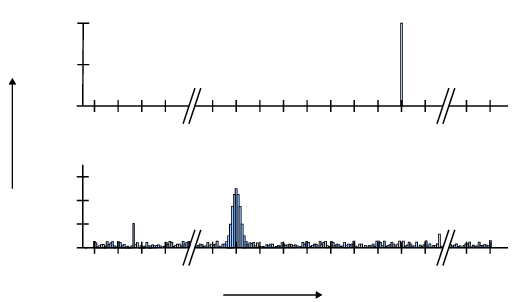}
    \caption{An example system at time $k$, where the cost-to-go happens to coincide with the disturbance as $J_{k+1}(w_{k})=w_{k}(u_k)$, and its dynamics are simply $x_{k+1}=w_k(u_k)$, with $u_k\in\{0,1\}$. The distributions $r(w_k|u_k)$ over $w_k\in\mathcal{W}=\{0,1,\dots,10000\}$, as a function of $u_k$ are obtained from (noisy) data and displayed in the figure. An engineer who aims to control this system in a semi-robust way using the KL-regularized framework is encouraged to reduce $\gamma_{E}$.}
    \label{fig:down}
\end{figure}

In the example scenario of Fig.~\ref{fig:down}, KL-regularized robust control (diagonal $\gamma_{E}=\gamma_{H}$, see Fig.~\ref{fig:map}) will judge the expected cost-to-go associated with $u_k=1$ between $10000$ for $\gamma_{E}=\gamma_{H}=0$ (worst-case), and $\sim \!\! 4998$ for $\gamma_{E}=\gamma_{H}\to\infty$ (expected value). It further sees the cost of $u_k=0$ (correctly) as $4000$ and will thus prefer this input. The engineer may interpret the uniform noise floor of the disturbance profile of $u_k=1$ as either spurious or otherwise irrelevant, and its expected cost closer to $2000$. In such a scenario, an engineer using our \emph{minsoftmax} approach can move into the $\gamma_{E}<\gamma_{H}$ interior to bias the adversary away from unlikely disturbances and achieve performance improvements on the underlying system.

\begin{figure}[t]
    \centering
    \input{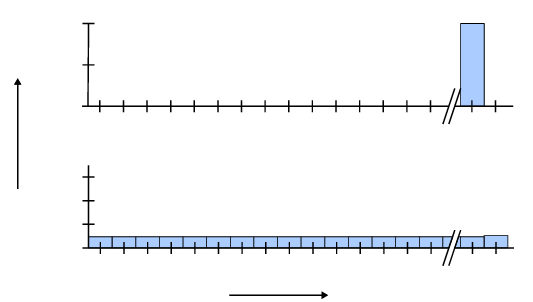}
    \caption{A second example system at time $k$, where the cost-to-go also happens to coincide with the disturbance as $J_{k+1}(w_{k})=w_k$, and its dynamics are state independent $x_{k+1}=w_k(u_k)$. This time, the distribution $r$ over $w_k\in\mathcal{W}=\{0,1,\dots,10000\}$ displayed in the figure is interpreted as an a-priori known distribution of the disturbance. An engineer who aims to control this system in a semi-robust way using the $\hinf$-framework is encouraged to increase $\gamma_{E}$.}
    \label{fig:up}
\end{figure}

In contrast, in a scenario like Fig.~\ref{fig:up}, pure $\hinf$-control ($\gamma_{E}=0$, $\gamma_{H}>0$, see Fig.~\ref{fig:map}) will judge the cost of $u_k=1$ as $10000$, independent of your choice of $\gamma_{H}\in[0,\infty)$, as disturbance $w_k=10000$ is both the worst-case and most likely. Tuning $\gamma_{E}>0$ in this scenario ensures the controller stops being ``blind'' to the additionally available stochastic information and soon starts preferring $u_k=1$ over $u_k=0$ (which in fact has a $\sim 99.99$\% chance of being the superior decision at any time).

\def\svgwidth{0.6\textwidth} 

\def\svgwidth{0.6\textwidth} 

\section{Conclusions and future work}\label{sec:6}

We have introduced the minsoftmax approach to robust controller design. The method is shown to enable controller synthesis that yields a combination of control-theoretical features in a single controller. We have shown how the inner problem of our regularized robust formulation can be solved analytically and use this to obtain a greatly simplified solution algorithm.

Possible future research directions include extending the minsoftmax approach to estimation. Furthermore, we expect that the method is compatible with a larger class of continuous space problems than those discussed here. The connection between a continuous space problem and its finite space abstraction also remains an open question.

\section{Acknowledgements}

We would further like to thank \emph{Dr.\ Giannis Delimpaltadakis} for our many interesting discussions on this work.

\section{Technical Proofs}\label{sec:technical_results}

\textit{Proof of Theorem \ref{thm:analytical_solution_continuous}} Let us first rewrite our candidate for optimality~\eqref{eq:softmax_optimal_adversary} explicitly as
\begin{equation}
p(w)^{*}=\frac{r(w)^{\gamma_{H}/\gamma_{E}}e^{J(w)/\gamma_{E}}}{\int_{\mathbb{R}^{n_{\text{w}}}}r(\hat{w})^{\gamma_{H}/\gamma_{E}}e^{J(\hat{w})/\gamma_{E}}~\mathrm{d}\hat{w}}. \label{eq:maximizer_explicitly}
\end{equation}
assuming integrals are well defined.
Substituting this into our regularization terms yields
\[
\begin{split}
    -\gamma_{H}&H_{\text{c}}(p^{*},r)+\gamma_{E}H(p^{*}) \\
    &=\int_{\mathbb{R}^{n_{\text{w}}}}p^{*}(w)\left(\gamma_{H}\log r(w) -\gamma_{E}\log p^{*}(w)\right)~\mathrm{d}w, \\
    &=-\gamma_{E}\int_{\mathbb{R}^{n_{\text{w}}}}p^{*}(w)\log\frac{p^{*}(w)}{r(w)^{\gamma_{H}/\gamma_{E}}}~\mathrm{d}w,
\end{split}
\]
    in which we substitute~\eqref{eq:maximizer_explicitly} to yield
\[
\begin{split}
    &=-\gamma_{E}\underset{\mathbb{R}^{n_{\text{w}}}}{\int}p^{*}(w)\log\frac{e^{J(w)/\gamma_{E}}}{\int_{\mathbb{R}^{n_{\text{w}}}}r(\hat{w})^{\gamma_{H}/\gamma_{E}}e^{J(\hat{w})/\gamma_{E}}~\mathrm{d}\hat{w}}~\mathrm{d}w, \\
    &=-\gamma_{E}\int_{\mathbb{R}^{n_{\text{w}}}}p^{*}(w)\Big(\log e^{J(w)/\gamma_{E}} \\
    &\hspace{2cm}-\log\int_{\mathbb{R}^{n_{\text{w}}}}r(\hat{w})^{\gamma_{H}/\gamma_{E}}e^{J(\hat{w})/\gamma_{E}}~\mathrm{d}\hat{w}\Big)~\mathrm{d}w, \\
    &=\gamma_{E}\log\int_{\mathbb{R}^{n_{\text{w}}}}e^{\alpha(w)/\gamma_{E}}~\mathrm{d}w-\underset{w\sim p^{*}}{\mathbb{E}}J(w), \\
\end{split}
\]
where $\alpha(w)=\gamma_{H}\log r(w)+J(w)$, which, through substitution, yields
\begin{equation}
\begin{split}
    -\gamma_{H}H_{\text{c}}(p^{*},r)+\gamma_{E} H(p^{*})+&\underset{w\sim p^{*}}{\mathbb{E}}J(w) \\
    &=\gamma_{E}\log\int_{\mathbb{R}^{n_{\text{w}}}}e^{\alpha(w)/\gamma_{E}}~\mathrm{d}w,\label{eq:continuous_optimal_cost}
\end{split}
\end{equation}
which is the cost at $p=p^{*}$.

We conclude our proof by showing the optimality of $p^{*}$ through the proof that inequality
\begin{equation}
\begin{split}
-\gamma_{H}H_{\text{c}}&(p,r)+\gamma_{E}H(p)+\underset{w\sim p}{\mathbb{E}}J(w) \\
&\le -\gamma_{H}H_{\text{c}}(p^{*},r)+\gamma_{E}H(p^{*})+\underset{w\sim p^{*}}{\mathbb{E}}J(w),\label{eq:inequality_proof_optimality}
\end{split}
\end{equation}
holds for all $p\in\mathcal{P}^{n}$.
We again rewrite our regularization terms as follows
\[
\begin{split}
-&\gamma_{H}H_{\text{c}}(p,r)+\gamma_{E}H(p) \\
&=-\gamma_{E}\int_{\mathbb{R}^{n_{\text{w}}}}p(w)\log\frac{p(w)}{r(w)^{\gamma_{H}/\gamma_{E}}}~\mathrm{d}w,\\
&=-\gamma_{E}\Big( \int_{\mathbb{R}^{n_{\text{w}}}}p(w)\log\frac{p(w)}{p^{*}(w)}~\mathrm{d}w \\
&\hspace{2cm}+ \int_{\mathbb{R}^{n_{\text{w}}}}p(w)\log\frac{p^{*}(w)}{r(w)^{\gamma_{H}/\gamma_{E}}}~\mathrm{d}w \Big),
\end{split}
\]
in which we again substitute~\eqref{eq:maximizer_explicitly} to yield
\[
\begin{split}
&=-\gamma_{E}\Big( \operatorname{KL}(p\|p^{*})\\
&\hspace{0.8cm} +\int_{\mathbb{R}^{n_{\text{w}}}}p(w)\log\frac{e^{J(w)/\gamma_{E}}}{\int_{\mathbb{R}^{n_{\text{w}}}}r(\hat{w})^{\gamma_{H}/\gamma_{E}}e^{J(\hat{w})/\gamma_{E}}~\mathrm{d}\hat{w}}~\mathrm{d}w \Big), \\
&=-\gamma_{E}\Big( \operatorname{KL}(p\|p^{*}) + \frac{1}{\gamma_{E}}\underset{w\sim p}{\mathbb{E}}J(w) \\
&\hspace{3.5cm}- \log\int_{\mathbb{R}^{n_{\text{w}}}}e^{\alpha(w)/\gamma_{E}}~\mathrm{d}w\Big), \\
\end{split}
\]
which we may substitute into the left-hand side of~\eqref{eq:inequality_proof_optimality} to obtain
\[
\begin{split}
-\gamma_{H}&H_{\text{c}}(p,r)+\gamma_{E}H(p)+\underset{w\sim p}{\mathbb{E}}J(w)= \\
&-\gamma_{E} \operatorname{KL}(p\|p^{*}) + \gamma_{E}\log\int_{\mathbb{R}^{n_{\text{w}}}}e^{\alpha(w)/\gamma_{E}}~\mathrm{d}w, 
\end{split}
\]
which, together with the substitution of~\eqref{eq:continuous_optimal_cost} into the right-hand side of~\eqref{eq:inequality_proof_optimality}, simplifies the inequality to
\[
\begin{split}
-\gamma_{E} \operatorname{KL}(p\|p^{*}) + &\gamma_{E}\log\int_{\mathbb{R}^{n_{\text{w}}}}e^{\alpha(w)/\gamma_{E}}~\mathrm{d}w \\
\le &\gamma_{E}\log\int_{\mathbb{R}^{n_{\text{w}}}}e^{\alpha(w)/\gamma_{E}}~\mathrm{d}w, 
\end{split}
\]
the validity of which is obvious as $\operatorname{KL}(p\|p^{*})>0$ for all $p\ne p^{*}$. 
\qed

\vspace{2mm}

\textit{Proof of Theorem \ref{thm:continous_recursion}}
Assume $\gamma_E >0$ and that Assumptions~\ref{as:1}, \ref{as:2} are satisfied.
Moreover, let $\xi_k = Ax_k + Bu_k$ be the system dynamics without disturbance. We then write~\eqref{eq:alpha}, using $r(w|x_k,u_k) = 
\rho e^{-\frac{1}{2}w^{\top}w}$, where $\rho = \frac{1}{\sqrt{2\pi}^{n_{\text{w}}}}$, as 
\[
\begin{split}
&\alpha(x_k,u_k,w_k) = \gamma_H\log \rho -\frac{\gamma_H}{2}w^{\top}w +\xi_k^\top P_{k+1} \xi_k \\
&\hspace{1cm}+ 2\xi_k P_{k+1} D^\top w + w^\top (D^\top P_{k+1} D) w + \zeta_{k+1},
\end{split}
\]
We can then write the solution to the inner optimization problem according to Theorem \ref{thm:analytical_solution_continuous} as 
\[
G(x_k,u_k) = \gamma_{E} \log \int_{\mathbb{R}^{n_{\text{w}}}}e^{\alpha_k(x,u,w)/\gamma_{E}}~\mathrm{d}w.
\]
Hence, the DP recursion at time $k$ can be written as 
\[
    J(x_k) = \min_{u_k} x_k^\top Q x_k + u_k^\top R u_k + G(x_k,u_k).
\]
By factoring out the terms constant w.r.t.\ $w$, as $c_{\xi}=(\gamma_{H}\log \rho +\xi_k^{\top}P_{k+1}\xi_k+\zeta_{k+1})/\gamma_{E}$, we can rewrite the remaining exponent of the integral as 
\[
-\frac{1}{2}w^\top(\gamma_{H}I - 2D^\top P_{k+1} D) w/\gamma_{E} + 2\xi P_{k+1} D^\top w/\gamma_{E}.
\]
Define $M_{k+1} = \gamma_H I - 2D^\top P_{k+1} D$, and notice that we must have $M_{k+1} \succ 0$ to ensure the integral converges. Define shorthand $b_{\xi} = 2D P \xi$ and complete the squares in the exponent 
\[
\begin{split}
&-\frac{1}{2}w^\top M_{k+1} w + b_{\xi}^\top w =\\
&\hspace{1cm}-\frac{1}{2}(w - \mu)^\top M_{k+1} (w - \mu)  + \underbrace{\mu^\top M_{k+1} \mu}_{\text{constant}},
\end{split}
\]
where $\mu = M_{k+1}^{-1}b_{\xi}$. Take the constant $\mu^{\top}M_{k+1}\mu=b_{\xi}^\top M^{-1}_{k+1}b_{\xi}$, out of the integration and consult~\cite[Sec. 8]{Petersen2008} for the explicit formula of the integral which  evaluates to 
\[
\int_{\mathbb{R}^{n_w}}e^{-\frac{1}{2\gamma_E}(w - \mu)^\top M_{k+1} (w - \mu)} \mathrm{d}w =  (\gamma_E2\pi)^\frac{n_{\text{w}}}{2} \operatorname{det}(M_{k+1})^{-\frac{1}{2}}
\]
Pugging back the constants we took out gives
\[
\begin{split}
&G(x_k,u_k)=\\
&\gamma_E \log \left( e^{b_{\xi}^\top M^{-1}_{k+1}b_{\xi}/\gamma_{E} + c_{\xi}}\sqrt{\gamma_E2\pi}^{n_{\text{w}}}/\sqrt{\operatorname{det}(M_{k+1})}\right),
\end{split}
\]
which, after keeping the relevant part and taking the logarithm, yields 
\[
G(u_k,x_k) = \xi^\top P_{k+1} \xi + \xi^\top P_{k+1}  D M_{k+1}^{-1} D^\top P_{k+1} \xi + \zeta_k,
\]
for
\[
\begin{split}
&\zeta_k = (\gamma_{E}-\gamma_{H})\log (2\pi)^{n_{\text{w}}}/2\\
&\hspace{2cm}-\gamma_{E}\log(\det(M_{k+1})/\gamma_{E}^{n_{\text{w}}})/2 + \zeta_{k+1}.
\end{split}
\]
Now define $F_a(P_k) = P_k  +  P_k D M_k^{-1} D^\top P_k$,
hence
\[
G(x_k,u_k) =(Ax_k+Bu_k)^\top F_a(P_{k+1}) (Ax_k+Bu_k) + q_k.
\]
Set $\frac{\partial J}{ \partial u_k} = 2 R u_k + 2B^\top F_a(P_{k+1}) (Ax_k + Bu_k) = 0$, we recover the optimal control law 
\[
u_k^* = - (R + B^\top  F_a(P_{k+1}) B)^{-1} B^{\top}  F_a(P_{k+1})Ax_k,
\]
which is unique due to the Hessian of the objective function with respect to $u_k$, given 
by $R + B^\top F_a(P_{k+1})B$,  being positive definite under $R \succ 0,M_k \succ 0$ for all $k$.

Remember that $J(x_k) = x_k P_k x_k + \zeta_k$ to obtain
$$
P_k = Q + K^\top R K + (A-BK)^\top F_a(P_{k+1}) (A-BK), 
$$
which, after working out the expression, gives
\begin{align}
\begin{split}
&P_{k}=Q+A^{\top}F(P_{k+1})A\\
&-A^{\top}F(P_{k+1})B(R+B^{\top}F(P_{k+1})B)^{-1}B^{\top}F(P_{k+1})^{\top}A.
\end{split}
\end{align}
Initializing with $P_h = Q_h$, $\zeta_h = 0$ and applying the recursion, we recover the algorithm in Theorem~\ref{thm:continous_recursion}.
We can explicitly write down an optimal adversarial distribution recognizing the Gaussian-like shape for the inner maximization problem. The normalization constant is easily calculated, but irrelevant. We have
$$
p_k^*(x_k,u_k) = \mathcal{N} \left(M_{k+1}^{-1} 2 D P_{k+1} (Ax_k - Bu^*_k) , \gamma_E M_{k+1}^{-1} \right).
$$
\qed

\bibliography{IEEECL}
\bibliographystyle{IEEEtran}

\end{document}